\documentstyle[aps,prl,floats]{revtex}

\newcommand{\bastar}{\begin{eqnarray*}}
\newcommand{\eastar}{\end{eqnarray*}}
\newskip\humongous \humongous=0pt plus 1000pt minus 1000pt

\newif\ifdtup

\relax
\newcommand{\be}{\begin{equation}}
\newcommand{\ee}{\end{equation}}
\newcommand{\bea}{\begin{eqnarray}}
\newcommand{\eea}{\end{eqnarray}}
\newcommand{\X}{{\vec X}}
\newcommand{\pro}{\partial}
\newcommand{\n}{\hat n}
\newcommand{\oneg}{\displaystyle\frac{1}{g}}

\newcommand{\D}{{\hat D}}

\newcommand{\A}{{\vec A}}
\newcommand{\valpha}{{\vec \alpha}}

\newcommand{\dfrac}{\displaystyle\frac}
\newcommand{\ba}{\begin{array}}
\newcommand{\ea}{\end{array}}

\newcommand{\nn}{\nonumber}
\begin{document}
                         \twocolumn[\hsize\textwidth\columnwidth\hsize\csname@twocolumnfalse%
                         \endcsname
\title  {Effective Theory of QCD}
\bigskip

\author{Y. M. Cho$^{1,2}$, Haewon Lee$^{3}$, D. G. Pak$^{1,4}$}

\address{
$^{1)}$Asia Pacific Center for Theoretical Physics,
Seoul 130-012, Korea \\
$^{2)}$Department of Physics, College of Natural Sciencies, Seoul National University,
Seoul 151-742, Korea
\\$^{3)}$Department of Physics, Chungbuk National University, Cheongju, Chungbuk 361-763, Korea
\\ $^{4)}$Department of Theoretical Physics, Tashkent State University,
Tashkent 700 095, Uzbekistan\\
{\scriptsize \bf ymcho@yongmin.snu.ac.kr, hwlee@phys.chungbuk.ac.kr,
 dmipak@apctp.kaist.ac.kr} \\ \vskip 0.3cm
}
\maketitle

\begin{abstract}
 We derive a generalized Skyrme-Faddeev action as the effective action of QCD
 in the low energy limit.  Our result demonstrates the existence of a mass gap
 in QCD which triggers the confinement of color.

\vspace{0.3cm}
\end{abstract}

\narrowtext
%\widetext
\bigskip
                           ]
Recently Faddeev and Niemi have discovered the
 knot-like topological solitons in the Skyrme-type
non-linear sigma model \cite{faddeev1}, and conjectured that the Skyrme-Faddeev
Lagrangian could be interpreted as an effective
Lagrangian for QCD in the low energy limit \cite{faddeev2}. The purpose of this Letter is to
derive the effective action of QCD from the first principles.
{\it Based on the parametrization of the gluon potential which emphasizes its topological character,
we derive a generalized  Skyrme-Faddeev action as
the effective action of QCD in the infra-red limit}.
In doing so {\it we demonstrate the existence of a mass gap which
generates the dual Meissner effect and triggers the confinement of color in QCD}.
Our result establishes a deep connection which exists between the Skyrme-Faddeev type non-linear sigma model and
QCD, and
 strongly endorses the idea that the knot solitons of Faddeev-Niemi type
could play an important role in QCD.

It has long been argued that the confinement of color in QCD could
take place through the dual Meissner effect \cite{nambu,cho1}.
But a rigorous derivation of an effective theory of QCD which could
demonstrate this conjecture
has been missing so far \cite{cho2,ezawa}.  To obtain the desired effective theory of QCD
it is important to understand the mathematical structure of the
 non-Abelian gauge theory. A characteristic feature of the non-Abelian gauge theory
is its rich topological structure manifested by the non-Abelian monopoles \cite{wu,cho3},
the $\theta$ -vacuua and the instantons \cite{bpst,thooft}.
So it is crucial to make sure that these topological characters
are properly taken into account in the effective theory.
To be specific consider $SU(2)$.
Here the relevant topology is the $\pi_2 (S^2)$ which describes the non-Abelian monopoles,
and the $ \pi_3(S^3) \simeq\pi_3(S^2)$ which describes the multiple vacuua
and the corresponding vacuum tunneling by instantons. To take into account the above topology
one needs to introduce a $CP^1$ field, or equivalently an isotriplet unit vector field $\n$,
in the theory.
This can be done with a natural reparametrization of connection which contains the
topological degrees explicitly \cite{cho1,cho2}. With the reparametrization one can obtain the
effective theory of QCD by integrating out all the dynamical degrees of the non-Abelian gauge theory.

Consider $SU(2)$ for simplicity.  A natural way to accomodate the  topological degrees
into the theory is to decompose the connection into the ``Abelian'' part which leaves $\n$
a covariant constant, and the remaining part
which forms a covariant vector field \cite{cho1,cho2},
\bea
  \vec{A}_\mu &=&A_\mu \n - \oneg \n\times\pro_\mu\n+\X_\mu\nonumber
  \\          &=& \hat A_\mu + \X_\mu, \,\,\,\,\,\,  (\n^2 =1),
\eea
where $
A_\mu = \n\cdot \vec A_\mu$
is the ``electric'' potential.
Notice that the Abelian projection $\hat A_\mu$ is precisely the connection which
leaves $\n$ invariant under the parallel transport
\bea
\D_\mu \n = \pro_\mu \n + g {\hat A}_\mu \times \n = 0.
\eea
Under the infinitesimal gauge transformation
\bea
\delta \n = - \vec \alpha \times \n  \,,\,\,\,\,
\delta \A_\mu = \oneg  D_\mu \vec \alpha,
\eea
one has
\bea
&&\delta A_\mu = \oneg \n \cdot \pro_\mu \valpha,\,\,\,\
\delta \hat A_\mu = \oneg \D_\mu \valpha  ,  \nn \\
&&\hspace{1.2cm}\delta \X_\mu = - \valpha \times \X_\mu  .
\eea
Notice that $\hat A_\mu$ still describes an $SU(2)$ connection which enjoys  the full $SU(2)$ gauge degrees of freedom.

With (1) one has
\bea
\vec{F}_{\mu\nu}&=&\hat F_{\mu \nu} + \D _\mu \X_\nu - \D_\nu \X_\mu + g\X_\mu \times \X_\nu    \nn \\
&=&(F_{\mu\nu}+H_{\mu\nu} + X_{\mu \nu}  )\n +  \D _\mu \X_\nu - \D_\nu \X_\mu,
\eea
where
\bea
&\hat F_{\mu\nu} =  (F_{\mu\nu}+H_{\mu\nu})\n, \nn \\
&F_{\mu\nu}=\pro_\mu A_\nu-\pro_\nu A_\mu,\,\,\,
H_{\mu\nu}=-\oneg\n\cdot(\pro_\mu\n\times\pro_\nu\n ),  \nn \\
&X_{\mu \nu }= g \n \cdot (\X_\mu \times \X_\nu).
\eea
Notice that $H_{\mu \nu }$ with $\n\!=\!\hat r$ describes the Wu-Yang monopole \cite{wu,cho3}.
Since $\pro_\mu {\tilde H}_{\mu \nu}\!=\!0$ except for the isolated singularities of $\n$,
one can introduce the ``magnetic'' potential $C_\mu$ for
$H_{\mu\nu}$ locally sectionwise. Indeed with
$\n = (\sin \alpha \cos \beta , \sin \alpha  \sin \beta , \cos \alpha)$,
one finds
\bea
H_{\mu \nu } = \pro_\mu C_\nu - \pro _\nu C_\mu, \,\,
C_\mu  = \oneg \cos \alpha \pro_\mu \beta.
\eea
Clearly
$C_\mu $ describes the Dirac's monopole potential around the isolated singularities of $\n$ .

Now the Yang-Mills Lagrangian can be expressed in terms of ${\hat A}_\mu$ and $\X_\mu$,
\bea
{\cal L} = &-&\dfrac{1}{4} \vec F^2_{\mu \nu }
=-\dfrac{1}{4}
{\hat F}_{\mu\nu}^2 -\dfrac{g}{2} {\hat F}_{\mu\nu} \cdot (\X_\mu \times \X_\nu)  \nn \\
 &-&\dfrac{g^2}{4} (\X_\mu \times \X_\nu)^2  -  \dfrac{1}{4} ( \D_\mu \X_\nu - \D_\nu \X_\mu)^2 .
\eea
Clearly it
 can be viewed as the restricted gauge theory
made of the Abelian projection
\bea
{\cal L}_{(R)} = -\dfrac{1}{4} {\hat F}^2_{\mu\nu} = -\dfrac{1}{4}(F_{\mu\nu}+H_{\mu\nu})^2 ,
\eea
which has an additional gauge covariant charged vector field (the valence gluons) as its source. More importantly
the restricted gauge theory
describes the dual dynamics of QCD with the dynamical degrees of the Abelian subgroup $U(1)$ as the electric component and the
topological degrees of $SU(2)$ as the magnetic component \cite{cho1,cho2}.

It must be emphasized that in  the above parametrization \, {\it $\n$\, (or equivalently $C_\mu$)
represents the topological degrees of the non-Abelian theory which exist in addition to the six
dynamical degrees represented by the transverse components of $A_\mu$ and $\X_\mu$} \cite{cho2,cho4}.
In the conventional analysis the topological degrees have often been neglected because
one could always make $\n$ trivial and remove it from the theory by a gauge transformation,
at least sectionwise locally. But notice that, if one must include the topologically
non-trivial sectors into the theory, one can {\it not} neglect $\n$
because it can not be removed by a smooth gauge transformation.

With these preliminaries we are ready to derive the effective action of QCD.
In principle we can obtain the effective action expressed solely by the topological field $\n$ by
integrating out all the dynamical degrees of QCD. Consider the generating functional for (8)
\bea
W[J_\mu, {\vec J}_\mu]&=&
\int DA_\mu D \X_\mu  \exp [- i \int  ( \dfrac{1}{4} {\vec F}_{\mu\nu}^2  \nn \\
&+& A_\mu J_\mu +\X_\mu \cdot {\vec J}_\mu) d^4 x  ] .
\eea
In the functional integral we have to integrate out the six dynamical degrees
with a proper choice of a gauge, leaving $\n$ as a background.
Here we will do it in two steps, first obtain the effective action
of the restricted gauge theory by integrating out $\X_{\mu}$, and  then obtain the effective action for the
topological field $\n$ by integrating out the remaining $A_{\mu}$.
Before we perform the integral, it is important to notice that there are two types of gauge transformations
one can consider, the active type  and the passive type.
This is because one could treat $\n$ either as a covariant multiplet or simply as
a fixed background. The first view provides an active type described by (3),
but the second view provides the following passive type
\bea
 \delta \n =0, \,\,\,\,\,
 \delta \A_\mu = \oneg D_\mu \valpha,
\eea
from which one has
\bea
&&\delta A_\mu =\oneg \n \cdot D_\mu \valpha, \,\,\,\,\,
\delta {\hat A}_\mu = \oneg (\n \cdot D_\mu \valpha ) \n,  \nn  \\
&&\hspace{0.8cm}\delta \X_\mu= \oneg [ D_\mu \valpha -
(\n \cdot D_\mu \valpha ) \n].
\eea
With this we now calculate the effective action in one-loop approximation.

To perform the $D\X_\mu$ integral we first fix the gauge
with the condition
\bea
&&{\vec F}_1 = ({\hat D}_\mu \X_\mu) - {\vec f}_1 = 0 , \nn \\
&&\hspace{0.1cm} {\cal L}_{gf_1}= - \dfrac{1}{2\kappa}({\hat D}_\mu \X_\mu)^2 .
\eea
To obtain the corresponding Faddeev-Popov determinant one must keep in mind that one has to keep
$A_\mu$ fixed to integrate out $\X_\mu$. This means that one could treat as if $\delta \X_\mu$ has
 the full $SU(2)$ freedom, even though it has no $\n$-component. So from (12)
the Faddeev-Popov determinant can be written as
\bea
&M^{ab}_1 = \dfrac{\delta { F}^a_1}{\delta \alpha^b}  \simeq
 \oneg ({\hat D}_\mu { D}_\mu)^{ab}  .
\eea
With this the generating functional takes the form with $\kappa=1$,
\bea
W[{\vec {J}}_\mu]&=&\int D\X_\mu  Det \|M_1\|  \exp \{ -{i \int  [
 \dfrac {1}{4}{\hat F}_{\mu \nu}^2}\nn \\
 &-&\dfrac {1}{2} \X_\nu {\hat{D}}_\mu {\hat{D}}_\mu \X_\nu
+g{\hat F}_{\mu \nu} \cdot(\X_\mu\times\X_\nu)  \nn \\
&+&\displaystyle{\frac{g^2}{4}} {(\X_\mu\times\X_\nu)}^2
 +\X_\mu \cdot {\vec J}_\mu] d^4x  \} .
 \eea
%where $J_\mu = \n \cdot {\vec J}_\mu$.
To remove the quartic term of $\X_\mu$ we introduce an auxiliary antisymmetric tensor field
${\chi}_{\mu \nu}$
and express
the quartic term
as
\bea
&&\exp{[\dfrac{i}{4} \int X_{\mu \nu }^2 d^4x]}\nn \\
&&\hspace{1cm}=\int D{\chi}_{\mu \nu} \exp {[\dfrac
{i}{4}\int({\chi}_{\mu\nu}^2
-2i{\chi}_{\mu \nu }X_{\mu \nu}) d^4x ]}.
\eea
The integration over $\X_\mu$ results in the functional determinant
\bea
Det^{-\frac{1}{2}} K_{\mu\nu}^{ab} &=& Det^{-\frac{1}{2}}
[g_{\mu \nu} \delta^{ab} \hat D \hat D  \nn \\
&-&2g(F_{\mu \nu}+H_{\mu \nu})
          \epsilon^{abc} \n^c
- i \chi_{\mu \nu}
\epsilon ^{abc} \n^c ],
\eea
so that the generating functional is given by
\bea
&&W[{\vec {J}}_\mu]=\int D \chi_{\mu \nu} Det \|M_1\| \, Det^{-\frac{1}{2}} \| K \|\nn\\
&&\exp {\{-i \int [\frac {1}{4}{\hat F}_{\mu \nu}^2-\frac{1}{4} { \chi_{\mu \nu}}^2
 +\dfrac{1}{2}{\vec J}_\mu K^{-1}_{\mu \nu } {\vec J}_\nu   ]d^4 x\} } .
\eea
The integration over the auxiliary antisymmetric field
can be performed for the trivial classical configurations $ \chi_{\mu \nu} = 0$.
The determinants can be calculated in one-loop approximation using the dimensional regularization,
and those involving only divergent parts are given by
\bea
 \ln Det \|M_1\| &=& i  \dfrac{g^2}{6(4\pi)^2} \cdot \dfrac{1}{\varepsilon}
 \int (F_{\mu \nu} + H_{\mu \nu })^2 d^4 x , \nn  \\
   \ln {Det}^{-{\frac{1}{2}}} \|K\| &=& i
 \dfrac{5g^2}{3(4\pi)^2} \cdot \dfrac {1}{\varepsilon}
 \int (F_{\mu \nu} + H_{\mu \nu })^2 d^4 x .
\eea
So the resulting effective Lagrangian for the restricted QCD is written as
\bea
{\cal L}_{(R)\, eff} =- \dfrac{z_1}{4} {\hat F}_{\mu \nu}^2,\nn\\
z_1 = 1-\dfrac{22g^2}{3(4\pi)^2} \cdot \dfrac{1}{\varepsilon}.\
\eea
This is precisely what one could have expected. Indeed with ${\hat A}_{\mu}$
as the background one can easily argue that the effective action of the
restricted gauge theory must have the above functional form,
 because it must be invariant under the gauge transformation (3)
of the background field ${\hat A}_{\mu}$.
So the only possible modification of the action from the $\X_\mu$
integral is the overall renormalization of the field strength
${\hat F}_{\mu\nu}$ given by (20).

Now
consider
\bea
W[J_\mu] = \int DA_\mu \exp {[-i\int(\dfrac{z_1}{4} {\hat F}_{\mu\nu}^2
+A_\mu J_\mu)d^4x]}.
\eea
We
have to integrate out the electric degrees $A_\mu$ from the restricted QCD to
obtain the effective action for $\n$. For this
it is crucial to remember that the restricted theory still has the full
$SU(2)$ gauge degrees of freedom, even though
it contains only the dynamical degrees of the Abelian subgroup.
This means that we have to fix the full $SU(2)$ gauge degrees of ${\hat A}_\mu$
to integrate out $A_\mu$.
Now one could fix the gauge with the condition
\bea
\pro_\mu {\hat A}_\mu =\pro_\mu A_\mu \n + A_\mu \pro_\mu \n
-\oneg \n \times {\pro}^2 \n = 0.
\eea
But here we will choose a simpler condition
\bea
&&\hspace{0.5cm}{F}_2 = \pro_\mu A_\mu \n -\oneg \n \times {\pro}^2 \n -{\vec f}_2 = 0,\nn \\
&&\hspace{0.2cm}{\cal L}_{gf_2}= -\dfrac{1}{2\lambda} [(\pro_\mu A_\mu)^2 +\dfrac{1}{g^2}
(\n \times\partial^2\n)^2].
\eea
With this condition one obtains the following Faddeev-Popov determinant
 using the active gauge transformation (4)
 \bea
M_2^{ab} = \dfrac{\delta {F_2^a}}{\delta \alpha^b} &=& \delta ^{ab}
\partial^2
+(\n^a \pro_\mu \n^b - 2\n^b \pro_\mu \n^a ) \pro _\mu   \nn \\
& &+\n^a {\pro}^2 \n^b - \n^b {\pro}^2 \n^a .
\eea
One can calculate the determinant in one-loop approximation
\bea
\ln Det \|M_2\| &=& i \int \{ -\dfrac{{\mu_0}^2}{4\pi}\cdot
\dfrac{1}{\varepsilon} (\pro_{\mu} \n)^2\nn \\
&&\hspace{-1.5cm}+\dfrac{1}{6(4\pi)^2} \cdot \dfrac{1}{\varepsilon} [g^2 H_{\mu \nu}^2
-3(\pro_{\mu} \n  \cdot \pro_{\nu} \n)^2]\}d^4 x,\
\eea
where $\mu_0$ is a mass scale. Now one can integrate 
$A_\mu$ from  (21) with (23) and (25),
and  with $\lambda =1$ we obtain the following Lagrangian,
\bea
{\cal L}_{eff}&=&-\dfrac{{\mu_0}^2}{4\pi}\cdot \dfrac{1}{\varepsilon} (\pro_{\mu}\n)^2
-\dfrac{1}{4}(1-\dfrac{8 g^2}{(4\pi)^2}\cdot \dfrac{1}{\varepsilon})H_{\mu\nu}^2\nn\\
&-&\dfrac{1}{2(4\pi)^2}\cdot\dfrac{1}{\varepsilon}(\pro_{\mu}\n\cdot\pro_{\nu}\n)^2
-\dfrac{1}{2g^2}(\n\times \pro^2 \n)^2.
\eea
So after a proper renormalization the final effective Lagrangian
can be written as
\bea
{\cal L}_{eff} &=&-\dfrac{\mu^2}{2}(\pro_{\mu}\n)^2
-\dfrac{1}{4}(\pro_\mu\n\times\pro_\nu\n)^2\nn\\
&-&\dfrac{\alpha_1}{4} (\pro_{\mu} \n  \cdot \pro_{\nu} \n)^2
-\dfrac{\alpha_2}{2}(\n\times \pro^2 \n)^2,
\eea
where $\mu$, $\alpha_1$, and $\alpha_2$ are the renormalized 
coulpling constants.
This is nothing but a generalized Skyrme-Faddeev Lagrangian. This completes
the derivation of the effective action of QCD from the first principles.

A few comments are in order now. First, we have integrated out
$\vec{X}_\mu$ and $A_\mu$ separately in two steps, to emphasize the
importance of the restricted QCD.  But certainly one could integrate out them
simultaneously in one step, to obtain the desired effective action \cite{cho5}.
Secondly, it appears that we have over-exploited the gauge degrees
of freedom in the above calculation. This is {\it not} so.  With
the introduction of $\n$, one can actually view (3) and (11) as
independent, which together give us an extended $SU(2)\times SU(2)$
gauge degrees of freedom \cite{cho5}. Thirdly, one might certainly
be tempted to go further and decompose $\n$ into the classical and
quantum parts, and integrate out the quantum part of $\hat{n}$ to
obtain the effective action.  But we emphasize that it suffices to
integrate out only the dynamical degrees of QCD for us to obtain
the desired effective action \cite{cho5}. Finally, the effective
action
 that we have derived is not exactly the Skyrme-Faddeev action. The Skyrme-Faddeev action
is unique in the sense that it consists of all infra-red relevant and marginal (and Lorentz invariant) interactions of $\n$ which are at most
quadratic in time-derivative, which is necessary for a Hamiltonian interpretation of the theory [2].
Our effective action contains the extra interactions which are not quadratic in time-derivative.
 Whether this should cause any serious problem or not is not clear at this moment,
 which need to be studied further \cite{cho5}.

The most important feature of our analysis is the appearance of the kinetic term of the topological field $\n$. This
teaches us the following lessons. First this tells that the topological field $\n$ can be treated as a dynamical (i.e., propagating) field
after the quantum correction. Remember that in the parametrization of the gauge connection (1) one can {\it not} treat $\n$ as a
 dynamical field,
because it simply represents the topological singularities of the point-like non-Abelian monopole
 at the classical level \cite{cho4}.
With the quantum correction, of course, one should
count (not only the dynamical degrees but
also) the topological degrees  as the true physical degrees
of the non-Abelian gauge theory. Secondly and perhaps more importantly this demonstrates the existence of a mass gap in QCD.
In fact from the dimensional analysis the kinetic term automatically introduces a mass scale to QCD
which could naturally be interpreted as the confinement scale.
 The generation of the mass scale comes from the dynamical symmetry breaking, which should induce the magnetic
condensation and the dual Meissner
effect responsible for the confinement of color in QCD \cite{cho1,cho2}.

We conclude with the following remarks:\\
1) With the effective action at hand one could verify the conjecture
that the knot solitons could describe the glueball
states of QCD explicitly. For this one must first ask whether
the effective action of QCD allows the knot solutions of
Faddeev-Niemi type. The answer, most probably, should be in the affirmative.
If so, one could really try to test
experimentally whether QCD contains the topological knot
solitons as the glueball states.\\
 2) Our analysis confirms that it is the restricted QCD which plays the central
 role in the confinement of color \cite{cho1,cho4}. It has all
the non-Abelian characters of QCD,
with the degenerate vacuua and the non-Abelian monopoles. With
the decomposition (1) the valence gluons $\X_\mu$ become a colored
sources which one can add to (or remove from) the theory at one's disposal, and
is not likely to play any important role in the confinement mechanism. \\
3) A generalization of the above analysis to an arbitrary group $G$ should be straightforward. The connection can be decomposed into the Abelian
part which consists of the dynamical degrees of the maximal Abelian subgroup $H$
as well as the topological
degrees of $G$, and the gauge covariant part which consists of the dynamical degrees of $G/H$. The Abelian projection
defines the restricted gauge theory which describes the dual
dynamics of $G$, and the covariant
vector field plays the role of the colored source of the restricted theory \cite{cho2,cho4}.
The effective action is obtained by integrating out all the dynamical degrees of the non-Abelian gauge theory.

A more detailed discussion, including the gauge dependence of the effective action and the generalization to $SU(3)$,
will be discussed in a forthcoming paper \cite{cho5}.

One of us (YMC) thanks Professor L. Faddeev and Professor C. N. Yang for the fruitful discussions.
The work is supported in part by Korean Science and Engineering
Foundation through Center for Theoretical Physics, and by Korea
Research Foundation through Project 1998-015-D0054.
                  
\end{document}